\documentclass[journal=macromolecules,manuscript=article]{achemso}

\usepackage[version=3]{mhchem} % Formula subscripts using \ce{}

\usepackage{graphicx}% Include figure files
\usepackage{dcolumn}% Align table columns on decimal point
\usepackage{bm}% bold math
\usepackage[normalem]{ulem}% Tian: allow \sout{} to realize a deleting line
\usepackage{xcolor}%
\usepackage{amsmath}% useful mathematical things like underbraces
\usepackage{verbatim}% comments
\usepackage{mathtools}
\usepackage{epstopdf}
\usepackage{float}
\usepackage{lineno}
\usepackage{caption}
\usepackage[utf8]{inputenc}
\usepackage[T1]{fontenc}
\usepackage{mathptmx}
\usepackage{etoolbox}
\usepackage{multirow}
\usepackage{amssymb}
\usepackage{epsfig}
\author{Jiting Tian}
\affiliation{Institute of Nuclear Physics and Chemistry, China Academy of Engineering Physics, 621999 Mianyang, China}
\alsoaffiliation{Univ. Grenoble-Alpes, CNRS, LIPhy, 38000 Grenoble, France}

\author{Jean-Louis Barrat}
\affiliation{Univ. Grenoble-Alpes, CNRS, LIPhy, 38000 Grenoble, France}
\email{jean-louis.barrat@univ-grenoble-alpes.fr}

\author{Walter Kob}
\affiliation{Department of Physics, University of Montpellier and CNRS, F-34095 Montpellier, France}
\email{walter.kob@umontpellier.fr}

\title[An \textsf{achemso} demo]
 {Influence of preparation and architecture on the elastic modulus of
polymer networks}

\abbreviations{IR,NMR,UV}
\keywords{American Chemical Society, \LaTeX}

\begin{document}

\begin{tocentry}

Some journals require a graphical entry for the Table of Contents.
This should be laid out ``print ready'' so that the sizing of the
text is correct.

Inside the \texttt{tocentry} environment, the font used is Helvetica
8\,pt, as required by \emph{Journal of the American Chemical
Society}.

The surrounding frame is 9\,cm by 3.5\,cm, which is the maximum
permitted for \emph{Journal of the American Chemical Society}
graphical table of content entries. The box will not resize if the
content is too big: instead it will overflow the edge of the box.

This box and the associated title will always be printed on a
separate page at the end of the document.

\end{tocentry}

\begin{abstract}

The elastic modulus $G$ of a polymer network depends notably on parameters such as the initial concentration of the monomers before the synthesis ($\rho_0$), the density of the cross-linker, or the topology of the network. Understanding how these factors influence the stiffness of the sample is hampered by the fact that in experiments it is difficult to tune them individually. Here we use coarse-grained molecular dynamics simulations to study how these quantities, as well as excluded volume interactions, affect the elastic modulus of the network. We find that for a regular diamond network, $G$ is independent of the initial monomer concentration, while for disordered networks (monodisperse or polydisperse) the modulus increases with $\rho_0$, at odds with the classical predictions for rubber elasticity. Analysis of the network structure reveals that, for the disordered networks, defects contribute only weakly to the observed increase, and that instead the $\rho_0$-dependence of $G$ can be rationalized by the presence of a pre-strain in the sample. This pre-strain can be quantified by the topological factor introduced in the affine network theory (ANT). Comparison of  the disordered networks with their phantom counterparts reveals that weakly crosslinked systems show a stronger $\rho_0$-dependence of $G$ due to an increase in entanglements at higher $\rho_0$, and that the polydisperse networks contain more entanglements than the monodisperse ones with the same average strand length. Finally we discuss the quantitative application of ANT to the simulated real networks and their phantom counterparts and conclude that the presence of excluded volume effects must be comprehensively taken into account for reaching a qualitative understanding of the mechanical modulus of the disordered networks.

\end{abstract}

\section{Introduction}

Polymer networks, including gels and elastomers, are important in everyday life as well as advanced technology developments\cite{Creton2017, Gu2019}, with applications ranging from automotive tires, membranes, drug delivery\cite{hoare2008hydrogels}, tissue engineering\cite{lee2001hydrogels}, stretchable electronics\cite{rogers2010materials}, to food manufacturing\cite{vilgis2015soft}. One outstanding feature of these materials is that their mechanical properties can be tuned to meet a specific request of a given application, making them highly attractive for materials science. For example, the elastic shear modulus $G$, one of the most basic and important properties of polymer networks, can be varied by orders of magnitude, spanning the range from 1~Pa to 1~MPa. The widespread presence of polymer networks in applications has motivated many studies that attempt to connect their macroscopic properties with the microscopic ones (polymer concentration \cite{Akagi2013, Nishi2017}, junction functionality\cite{Wang2017}, connectivity of the network \cite{Wang2017}, chain polydispersity\cite{Sorichetti2021}) or the details on how the sample is prepared such as the presence of a solvent\cite{Yoshikawa2021, Tang2023, hagita2023all}, the density of defects\cite{Zhong2016, Lang2018, Panyukov2019, Lin2019}, or the process parameters~\cite{Gu2019}. Despite these efforts, we are at present still lacking a satisfying understanding on how the mechanical properties of polymer networks depend on the mentioned parameters, since their complex hierarchical and multiscale structure makes it difficult to advance on this question\cite{Danielsen2021}.

Experiments using end functionalized, monodisperse star polymers \cite{Akagi2013, Nishi2017} have clearly demonstrated that an increased initial concentration of the solution results in a higher modulus of the percolated monodisperse network. However, the mechanism that gives rise to this behavior has not yet been clarified. For the regime below the overlap concentration where defects (molecular loops and dangling ends) are easily formed  due to the large inter-molecular distances, some experimental works\cite{Akagi2013, Nishi2017} indicate that a network prepared at a lower concentration has more molecular defects and thus a smaller modulus, a result which is supported by several other studies\cite{Zhong2016, Lang2018, Panyukov2019, Lin2019}. However, for the regime beyond the overlap concentration, defects (especially dangling ends) rarely form while the increase of the modulus with concentration is still observed\cite{Akagi2013}, implying that a different mechanism is responsible for the experimental observations.

One important difficulty in experimental studies is that the modification of a single parameter (density, crosslinker concentration, etc.) can simultaneously affect multiple structural quantities of the network. For example, an increasing concentration of precursur polymers can concurrently lead to a lower density of defects\cite{Zhong2016}, more chain entanglements\cite{Schlogl2014}, or a weaker pre-strain\cite{Beech2023}, all of which influence the elastic modulus, and in an experiment it is very difficult to disentangle the contributions of these factors. In this respect, computer simulations\cite{Allen2017,frenkel2023understanding,tuckerman2023statistical} have a big advantage since they allow to modify one quantity systematically while keeping the other quantities unchanged, therefore providing insights into the relative importance and microscopic details of individual mechanisms. 

In this work, we use molecular dynamics simulations to prepare and deform coarse-grained models of polymer networks to reproduce the experimentally-observed concentration dependence of the elastic modulus, and analyze the network structures to uncover the relevant mechanisms that are responsible for the observed effects. We find that while defects and entanglements indeed contribute to the concentration dependence, pre-strain plays the most important role in this phenomenon. This conclusion holds not only for monodisperse networks generated by reactions of star polymers, but also for randomly-crosslinked polydisperse networks, therefore demonstrating the universal role of the pre-strain on the elasticity of polymer networks.

\section{Methods}

In this section we describe how we prepare three types of elastomer networks with various initial particle densities, deform them to get their shear moduli, and determine the pre-strain of a network.

\subsection{Preparation of coarse-grained polymer networks}

\subsubsection{Interaction potentials and general parameters}

We simulate coarse-grained polymer networks based on the well-established Kremer-Grest bead-spring model\cite{Kremer1990} using the LAMMPS software\cite{Plimpton1995}. Monomers (bonded to one or two particles) and crosslinkers (bonded to three or four particles) are represented by particles (beads) which have the same mass $m$ and are interacting through a Weeks-Chandler-Andersen (WCA) short range pair potential\cite{weeks1971role}, $i.e.$, the purely repulsive part of the Lennard-Jones potential:
\begin{equation}
 U_{\rm WCA}(r)=\begin{cases}
 4\epsilon \left[(\frac{\sigma}{r})^{12}-(\frac{\sigma}{r})^{6}\right]+\epsilon, & {\rm if}\ r<2^{1/6}\sigma, \\
 0, & \rm else,
 \end{cases}
\label{eq:1}
\end{equation}

\noindent
where $\sigma$ is the particle diameter and $\epsilon$ is the well depth of the Lennard-Jones potential. Chemically bonded beads are also interacting via a finite extensible nonlinear elastic (FENE) potential:

\begin{equation}
 U_{\rm FENE}(r)=-\frac{kr_0^2}{2}{\rm ln}\left[1-\left(\frac{r}{r_0}\right)^2\right],
\label{eq:2}
\end{equation}

\noindent
where $k=30\ \epsilon/\sigma^2$ and $r_0=1.5\ \sigma$. Mass, length, and energy will be expressed in units of $m$, $\sigma$, and $\epsilon$, respectively. Other units are derived from the three basic ones: Temperature $[T]=\epsilon/k_{\rm B}$, density $[\rho]=\sigma^{-3}$, time $[t]=\tau=\sqrt{m\sigma^2/\epsilon}$, and elastic moduli $[G]=\epsilon\sigma^{-3}$. If not mentioned otherwise, all simulations are performed at $T=1$ with periodic boundary conditions applied in all dimensions of the box. Depending on the parameters of the simulations we have typically $N_{\rm tot}\simeq 10^5$ particles in the simulation box. The time step of all simulations is set to 0.005 and we have verified that a smaller time step (0.003) gives statistically identical results. 

To avoid the difficulties of simulating the complex solvent effects on chain dynamics\cite{Zheng2021} and on network elasticity\cite{Yoshikawa2021, hagita2023all, Tang2023}, we consider here elastomers rather than gels, since the former contain no solvents. In particular we investigate how the architecture of the network and the particle density $\rho_0$, defined as $N_{\rm tot}$ divided by the initial simulation box volume, influence the mechanical properties of the system. To mimic the de-swelling step in the elastomer synthesis, we generate the networks (see below for details) in a cubic box with a given initial size (corresponding to $\rho_0=0.2$, 0.4, and 0.8), and then slowly compress the box at constant temperature to the same final target size thus giving the same final density $\rho=0.8$ (close to the melt density $\rho_{\rm melt}=0.85)$. Subsequently we relax each network under $NVT$ (constant volume and temperature) conditions for $2\times10^4\ \tau$, and calculate the average pressure $P_0$ from the instantaneous pressure values sampled during the second half of this simulation. (Typical values of $P_0$ are between 3.5 and 3.8, see Tab. S1 in the supporting information (SI).) Then the box is relaxed under $NPT$ conditions for $10^4\ \tau$ with $P_0$ independently applied to each dimension, which results in a box that is slightly non-cubic, but with an insignificant volume change. (Typical deviations from cubic shape are around 2\% while the largest deviation is about 5\%.) We have verified that $10^4\ \tau$ is sufficiently long to equilibrate the configurations since increasing the relaxation time to $10^5\ \tau$ does not change the results (not shown here). This $NPT$-relaxation simulation allows to remove the pre-existing internal stress, and hence avoids the necessity of using a more complicated form (Eq. (14) in Ref.\citenum{Sorichetti2021}) to fit the stress-strain data points. The obtained slightly non-cubic samples are then used as starting configurations for the subsequent simulations of the elastic deformation.

\subsubsection{Protocols for generating three network architectures}

To study the influence of the geometry of the networks, we consider two types of monodisperse network and one type of polydisperse network. The first monodisperse network, called ``diamond'', is generated by first placing crosslinkers at the sites of a diamond lattice and then connecting each closest pair by a strand of $N$ monomers. In this way, each crosslinker connects four monomers and each monomer connects 2 particles (crosslinker or monomer). Subsequently an energy minimization and a long ($10^4\ \tau$) $NVT$ simulation is used to relax and equilibrate the configurations. The resulting diamond networks have a perfectly regular topology and no topological entanglements (see section S3 in the SI), so that they can serve as ideal reference configurations, although it would be very difficult to synthesize them in experiments.

The second type of monodisperse network, termed ``amorphous'', is inspired by the experimental synthesis of polyethylene-glycol-gels (PEG-gels)\cite{Akagi2013,Nishi2017,Sakai2008}. In practice, two types of star polymers having 4~arms (A$_4$ and B$_4$) are put into the box in equal numbers. At this stage A and B are just labels, since both polymers have identical interactions. Then we use a $NVT$ simulation to allow the molecules to diffuse for a long time to reach a randomly distributed state having no memory of the initial positions, during which we turn off the inter-chain WCA potential to speed up the diffusion. After sufficient mixing, we gradually add back the WCA potential and then quickly heat the system to $T=10$ in order to relax the structure, before using a linear cooling schedule to decrease $T$ slowly ($10^3\ \tau$) to $T=1$. This heating-and-cooling operation helps the system to reach a stable state that has no energetically unfavorable local particle packings. At this stage we turn the end monomers of the arms into reactive, which means that they can form a chemical bond with each other if two of them are close enough (less than $2^{1/6}\sigma$) and have not yet reacted. Note that, similar to the experimental situation, we only allow two end monomers from {\it different} type (A and B) star polymers to react. Each star polymer has one center particle (as a crosslinker) and four arms, each of which contains $N/2$ monomers, so after one reaction a strand with $N$ monomers is formed. We choose $N=10$ and $N=50$ to generate networks which represent, respectively, slightly and strongly entangled systems, as will be shown in the Results section and in the SI. Diamond networks are also created using the two $N$ values for the sake of comparison.

For the generation of the amorphous networks, we simulate the system for a long time ($10^5\ \tau$) in the $NVT$ ensemble which allows that about 95\% of the end-monomers become linked. At the end of this reaction stage, it is possible that a small number of unreacted arm ends are too far away from each other and thus will never react due to the rigidification of the network. This results in a few dangling ends forming network defects, but their concentration is negligible, as shown in Tab. S1 in the SI. An advantage of this ``A-B reaction'' protocol is that no closed (first-order) loops are formed, although higher order loops are still possible, and their densities depend on the preparation concentration, see Fig. S2 in the SI. We note that a recent study revealed that high order loops (especially the second-order) can affect the elastic moduli of gels\cite{Zhong2016}. However, in the SI we estimate that the loops in our simulations of elastomers result in at most an increase of 27\% in $G$ between $\rho_0=0.2$ and $\rho_0=0.8$. Hence we conclude that these loops are not the main reason for the observed strong concentration dependence of $G$ (see Fig. \ref{fig:2}), i.e., the main conclusions of the present paper are not influenced by the existence of these loops.

Although in experiments the concentration dependence of $G$ is studied for monodisperse PEG-gels, we check and analyze the universality of this dependence in polydisperse networks. To this end, we prepare nearly defect-free polydisperse networks using the protocol pioneered by Grest and Kremer\cite{Grest1990}, making, however, a slight modification, see the sketch in Fig. S1(a) and (b) in the SI. In detail, we first generate a nearly-equilibrium solution containing 100 independent polymer chains with 1000 monomers in each chain using the radical-like-polymerization method\cite{Perez2008}. To get a defect-free polydisperse network, each end monomer is first linked to the closest monomer belonging to a different chain by a FENE bond to avoid dangling ends. This step will be denoted as ``end-linking'', and it changes the 200 end monomers (having exactly one bond) into normal monomers (having exactly two bonds) and changes 200 normal monomers into crosslinkers (having three bonds).
Subsequently we form bonds between pairs of randomly selected monomers that belong to different chains, that are closer than $2^{1/6} \sigma$, and that are not yet crosslinkers. This step will be referred to as ``chain-linking'', and results that each chain-link generates two (connected) crosslinkers (each having three bonds). Note that with this process a junction created by a chain-link is usually made up by two neighboring crosslinkers, different from the above-generated monodisperse networks where a junction always consists of only one crosslinker. Besides, in our polydisperse networks some junctions may consist of only one crosslinker if they are generated by end-linking, and some may be a group of multiple (more than two) interconnected crosslinkers if two or more crosslinks (end-link or chain-link) happen to be adjacent, see an example in Fig. S1(c) in the SI. 

At the end of this procedure the system has a total of $p_{\rm c}N_{\rm tot}$ crosslinkers (generated via end-linking or chain-linking). The two crosslinker fraction values, $p_{\rm c}=0.018$ and $p_{\rm c}=0.098$, are chosen to get an average strand length $\left \langle N \right \rangle\simeq50$ and $\left \langle N \right \rangle\simeq10$, respectively, hence reproducing the values of the monodisperse networks. For simplicity, in the following we sometimes omit the brackets and directly use $N=X$ ($X=10$ or $X=50$) to indicate the polydisperse networks with $\left \langle N \right \rangle \simeq X$, while the reader should keep in mind that here $N$ refers to the average strand length. We will show in the Results section that the results for these polydisperse networks are consistent with those obtained for monodisperse networks.

Note that, different from the original Grest-Kremer protocol, Fig. S1(a), here we only allow two particles from {\it different} chains to link, Fig. S1(b), so that the formation of closed loops (mainly formed by self-chain linking) is considerably suppressed. Our modified protocol cannot completely avoid generating closed loops, as two chain-links may result in two connected crosslinkers on one chain and two chemically separated crosslinkers on another chain, thus forming a closed loop, see an example in Fig. S1(d). However, the number of these closed loops is negligible compared to that of normal strands (defined as a chain connecting two different junctions), see Tab. S1 in the SI. The effect of these closed loops and high-order ones is estimated by the recently-developed RENT\cite{Zhong2016} in the SI, showing that our main conclusions are not affected by the loops.

In total, we have prepared $3\times2\times3=18$ network configurations, including three architecture types (diamond, amorphous, and polydisperse), two $N$ values, and three $\rho_0$ values. Detailed information of each network is shown in Tab.~S1 in the SI. No matter what the network type is, our choice of $N_{\rm tot}\simeq 10^5$ leads to around 2000 and 8000 strands for $N=50$ and $N=10$, respectively, which is large enough to be self-averaging, so we do not need to generate multiple configurations for a given choice of $N$ and $\rho_0$.

\subsection{Elastic deformation of the networks}

Each network is deformed uniaxially from zero strain to an extension $\lambda=1.2$ by a succession of small elongations of size $\Delta\lambda=0.02$, each of which lasts a time $\Delta t$. After each strain increment the system is relaxed for a time $\Delta t$ by fixing the box size in the straining direction constant while the two other directions are kept at a pressure $P=P_0$. A configuration is saved after each elongation step, so at $\lambda=1.2$ one has saved for each deformation simulation a total of 10 deformed configurations that can subsequently be analyzed. The strain rate, defined by $\dot{\varepsilon}=0.02/(2\times\Delta t)$ and given by $\dot{\varepsilon}=2\times10^{-6}$ for $N=50$ and $\dot{\varepsilon}=2\times10^{-5}$ for $N=10$, is sufficiently small to ensure that the results are independent of this parameter, as decreasing it by a factor of two does not lead to significantly different results (see Fig. S6 in the SI).

Each saved configuration serves now as starting point of an independent relaxation run of $10^4\ \tau$ using the same dimension controls employed for the deformation run. From this run we collect, uniformly in time, 20,000 values of the instantaneous engineering stress, defined as $\sigma_{\alpha}=[\sigma_{\alpha\alpha}-0.5(\sigma_{\beta\beta}+\sigma_{\gamma\gamma})]/\lambda$, where $\alpha$ is the direction of deformation and $\sigma_{\alpha\alpha}$ is the $\alpha$-direction component of the virial stress without kinetic contributions, $i.e.$,

\begin{equation}
 \sigma _{\alpha \alpha }=-\frac{1}{V}\sum_{i}r_{i\alpha}f_{i\alpha} \quad .
\label{eq:3}
\end{equation}

\noindent
Here $V$ is the system volume, and $r_{i\alpha}$ and $f_{i\alpha}$ are, respectively, the $\alpha$-component of the position and force vectors of particle $i$. The collected $20,000$ values are then averaged to give a stress value for the elongation $\lambda$ along the direction $\alpha$. Although there are still some controversy on whether or not the virial stress should include kinetic terms~\cite{zhou2003new,subramaniyan2008continuum,shi2023perspective}, this is irrelevant here, since the kinetic energies are distributed equally among the three directions if the system is equilibrated and therefore the kinetic contribution to the deviatoric stress vanishes. We have in fact calculated the stress {\it including} the kinetic contributions and found no changes in the results, implying that our networks have been well equilibrated at each $\lambda$ during the deformation.

For each network, we perform the above deformation protocol along the three directions $x,y,z$ independently and average the obtained stress values for each $\lambda$. Finally the elastic modulus $G$ is fitted from the $\sigma(\lambda)$ data points using $\sigma=G(\lambda-1/\lambda^2)$, the classical stress-extension relation for polymer-network materials for which the elasticity is due to the entropy of the polymer chains~\cite{rubinstein2003polymer}. The stress-strain curves for all the 18 networks are shown in Fig. S7 in the SI, from which one sees that the quality of the fits is overall fairly good, and the obtained moduli are listed in Tab. S2 of the SI.

To explore the influence of the excluded volume and the entanglements,  we generate for each $NPT$-relaxed network its phantom counterpart by turning off the WCA potential and then relaxing it under $NVT$ conditions for $10^4\ \tau$. Subsequently we deform the network with the same protocol and parameters as used for the real networks, except that a $NVT$ rather than $NPT$ simulation is used, since the phantom networks would collapse under $NPT$ conditions due to the absence of excluded volume interactions. We obtain their moduli with the same fitting procedure and the fitted curves and the obtained moduli are presented in Fig. S8 and Tab. S2, respectively.

\subsection{Calculation of pre-strain}

The widely used rubber elasticity theories that are referred to as affine network model (ANM) and phantom network model (PNM)\cite{rubinstein2003polymer} assume that (1)~the network is made up of monodisperse phantom Gaussian strands (here ``Gaussian'' refers to a single chain's random-coil state which gives a harmonic force-extension relation), that (2)~the end-to-end vectors of the strands obey Gaussian statistics like in a polymer melt, that (3)~the time-averaged positions of junctions move in an affine way upon deformation, and that (4) the loops in the network are not entangled. Based on these four assumptions one can obtain a very simple expression for $G$ \cite{rubinstein2003polymer}:

\begin{equation}
G_{\rm ANM}=\nu_{\rm eff} k_{\rm B}T,
\label{eq:4}
\end{equation}

\noindent
for affinely-moving junctions, and

\begin{equation}
G_{\rm PNM}=\frac{f-2}{f}\nu_{\rm eff} k_{\rm B}T=0.5\nu_{\rm eff} k_{\rm B}T,
\label{eq:5}
\end{equation}

\noindent
for thermally-fluctuating junctions in a tree-like network, where $\nu_{\rm eff}$ is the effective strand density, $i.e.$, the number of normal (elastically-effective) strands divided by the box volume, and $f$ is the functionality (number of chemically bonded neighbors) of the junctions ($f=4$ in the present study so that the prefactor equals 0.5). Closed loops and dangling ends are elastically non-active, so they are not counted when calculating $\nu_{\rm eff}$. In the present study, we have chosen the preparation parameters so that networks with the same $N$ have very similar $\nu_{\rm eff}$ (see Tab.~S1 in the SI) and thus should give very similar moduli according to the above formula. However, below we will show that $G$ depends on the network type and $\rho_0$, therefore reflecting physics beyond these classical models.

In real networks it is difficult to satisfy simultaneously the aforementioned four assumptions. For example, experiments\cite{Beech2023} have shown that the strands in a network generated from a solution show some pre-strain due to the presence of quenched disorder when the system was generated, which is in line with results from simulations\cite{Gusev2019numerical,Gusev2019molecular,Tsimouri2020,Gusev2022} as well as our findings (details are given below), so that the second assumption, i.e., Gaussian statistics for the end-to-end vectors, usually does not hold except for networks generated from a melt. Gusev\cite{Gusev2019numerical} showed that if one follows the standard derivation of the classical models but does not apply this assumption, one arrives at the expression of the affine network theory (ANT),

\begin{equation}
G_{\rm ANT}=\Gamma\nu k_{\rm B}T,\ {\rm with}\ \Gamma =\left \langle \frac{\overline{\mathbf{R}} ^{2} }{N_{\rm b}b^2} \right \rangle,
\label{eq:6}
\end{equation}

\noindent
where $\overline{\mathbf{R}}$ is the time-averaged end-to-end vector of a strand and the bracket average is performed over all the strands including defects (dangling ends and loops), i.e., the ANT allows to take into account the influence of these defects\cite{Gusev2019numerical,Gusev2019molecular}. Note that here the strand density $\nu$ is calculated by counting all strands (including defects), different from the quantity $\nu_{\rm eff}$ used in the ANM and PNM. The quantity $\Gamma$ in Eq.~(\ref{eq:6})
measures how much the network is pre-stretched compared to the Gaussian (melt) state. Note that $N_{\rm b}b^2$ is the mean-squared end-to-end distance of an ideal chain with $N_{\rm b}=N+1$ being the number of bonds (recall that the strand chemical length $N$ is defined as the number of monomers) and $b=0.965$ is the bond length in the Kremer-Grest model.
 
Gusev and coworkers\cite{Gusev2019molecular} showed that, for the case of simulated phantom Gaussian networks, $\Gamma$ can be estimated using a maximum-entropy procedure and that the ANT predicts accurately the shear moduli of such networks\cite{Gusev2019numerical,Gusev2019molecular,Tsimouri2020}, although it fails to do so for real networks having excluded volume interactions~\cite{Tsimouri2020}. Following Gusev\cite{Gusev2022} we refer in the following to networks with excluded volume interactions as ``real networks'' to distinguish them from their phantom counterparts. Note that most derivations of the classical rubber models, i.e., ANM and PNM, assume that all the strands have the same chemical length $N$, while the ANT does not make this assumption, as one recognizes from the fact that $N_{\rm b}$ is {\it inside} the brackets in Eq. (\ref{eq:6}). As a result, the ANT can be expected to be applicable also to polydisperse networks as long as all the strands are sufficiently long to display upon stretching a Gaussian-extension behavior.

\begin{figure*}
\centering
\includegraphics[width=0.75\textwidth]{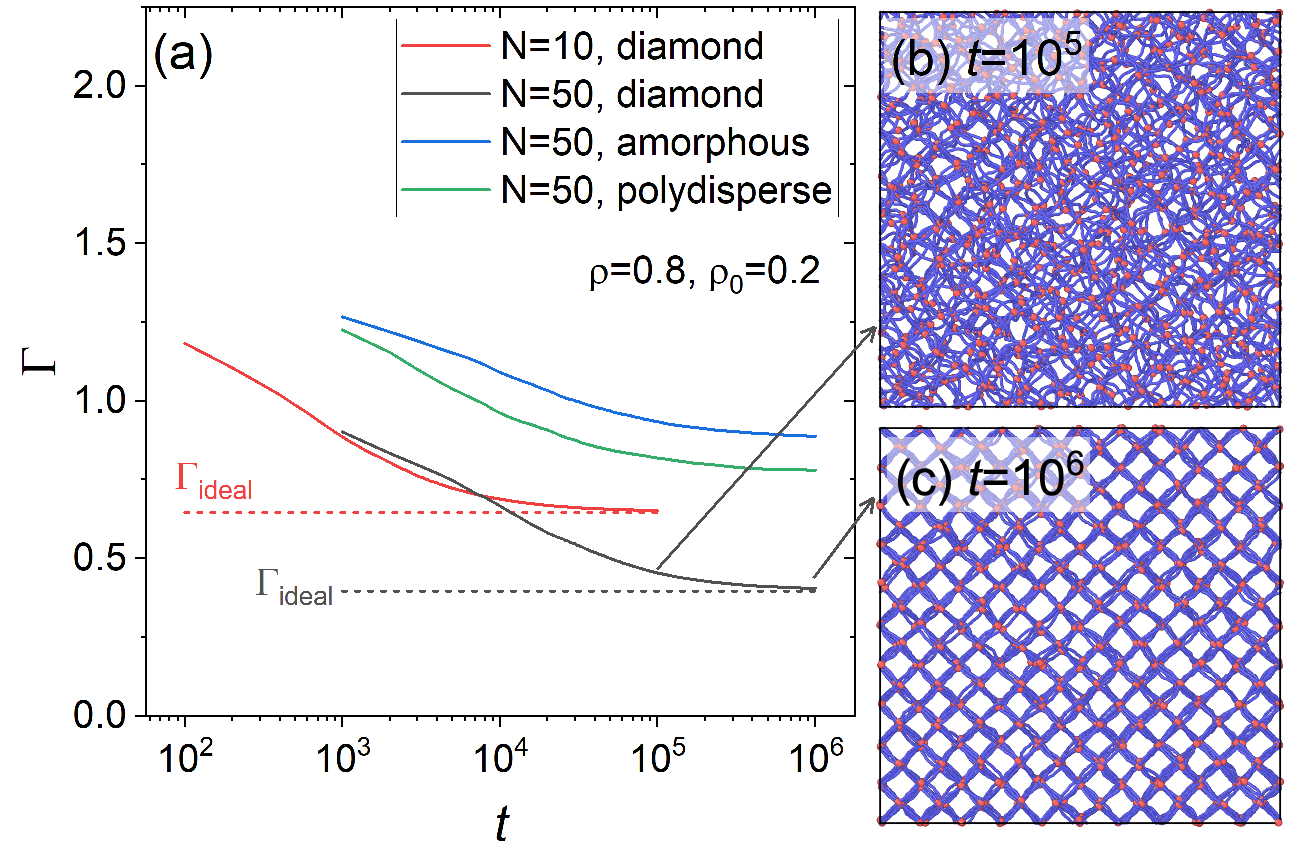}
\caption{(a) The pre-strain measure $\Gamma$ as a function of time $t$ for four typical networks with $\rho_0=0.2$. The dashed lines are the values of $\Gamma_{\rm ideal}$ calculated from the corresponding ideal diamond lattice. The two snapshots in (b) and (c) are the time-averaged configurations of $N=50$ diamond networks averaged up to $t=10^5$ (top) and $t=10^6$ (bottom), where one sees that the crosslinkers (red spheres) are indeed located approximately at the diamond lattice sites when $t=10^6$.
}
\label{fig:1}
\end{figure*}

In view of the success of the ANT for phantom Gaussian networks\cite{Gusev2019numerical,Gusev2019molecular,Tsimouri2020}, we use $\Gamma$ to characterize the pre-strain of our simulated real networks with various $\rho_0$, although a quantitative prediction of $G$ based on $\Gamma$ for real networks is at present not available. Since currently there are no efficient algorithms like the mentioned maximum-entropy method to estimate $\Gamma$ of real networks, we perform a very long ($10^5$ or $10^6$ $\tau$) $NVT$ simulation to sample $10^3$ instantaneous configurations to measure $\Gamma$. For a network that has the topology of a regular lattice, the value of $|\overline{\mathbf{R}}|$ is readily calculated and for our diamond lattice one finds $\left | \overline{\mathbf{R}} \right |=d_0$, where $d_0$ is the distance between two closest sites on the lattice, from which the value of $\Gamma$ can be immediately obtained. Comparison of the measured $\Gamma$ with the theoretical prediction allows thus to determine the time needed for the convergence of $\Gamma$. 

Figure \ref{fig:1}a presents the time evolution of $\Gamma$ for four different types of networks (undeformed)  with $\rho_0=0.2$. For the diamond network with $N=10$ one finds that $\Gamma$ converges to its theoretical value $\Gamma_{\rm ideal}=d_0^2/N_bb^2$ at around $t=10^5$, while for $N=50$ it takes $t=10^6$. The two snapshots in panels (b) and (c) show for the $N=50$ system the position of the particles after an average over $t=10^5$ and $t=10^6$, respectively, and one clearly sees how this average position takes the expected form of the diamond lattice at $t=10^6$. Also included in the graph are the data for the amorphous and polydisperse networks ($N=50$) and one recognizes that the time scale for convergence to a plateau is very similar to the one needed for the diamond network for $N=50$. In Fig. S9 we show that the converging speed of $\Gamma$ is independent of $\rho_0$.

We also mention that the mentioned time scales needed for the convergence of $\Gamma$ are significantly shortened if one puts the sample under stress. This is due to the fact that the chains are put under tension, i.e., they are less floppy. As a consequence the time scales for convergence of $\Gamma$ are of the order of $10^4\tau$.

We also calculate the pre-strain quantity of the initial networks before the compression (de-swelling), denoted as $\Gamma_0$. These initial networks have strands that are more stretched, thus making them stiffer than their final counterparts, leading to a much faster convergence of $\Gamma$, see Fig. S9(c). Finally we have also calculated $\Gamma$ of all the simulated phantom networks, using $t=10^5$, which has been verified to be sufficiently long for obtaining converged values. The so obtained ANT-related quantities of the networks are listed in Tab. S2.

\section{Results and discussion}

Here we first discuss the mechanical properties of the monodisperse networks, followed by the presentation of the results for the polydisperse networks. Finally we discuss to what extent the ANT is able to describe the elastic properties of these systems.

\subsection{Monodisperse networks}

\begin{figure*}[th]
\centering
\includegraphics[width=0.75\textwidth]{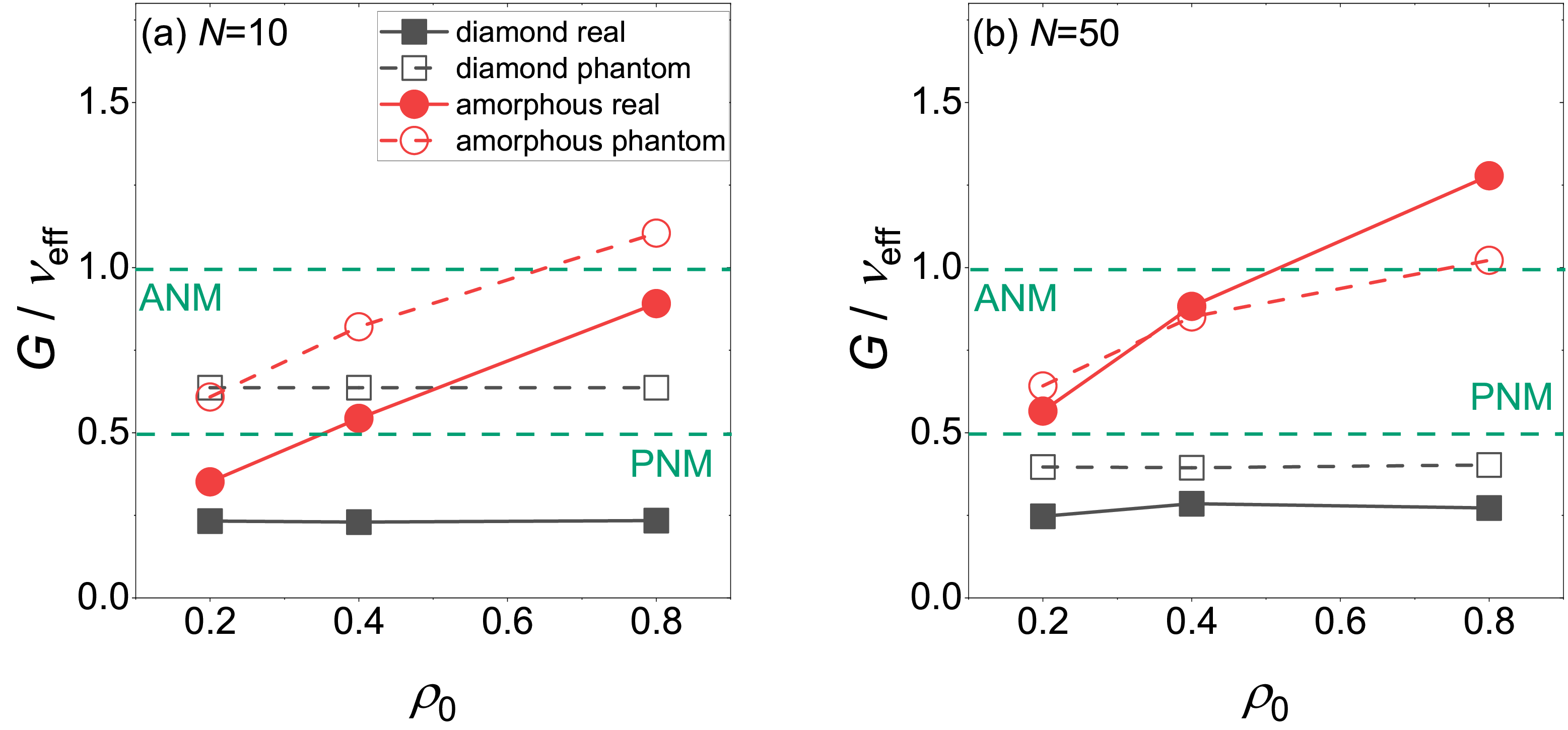}
\caption{Shear modulus $G$ (scaled by $\nu_{\rm eff}$) as a function of preparation concentration $\rho_0$ of simulated monodisperse networks with (a) $N=10$ and (b) $N=50$. The upper and lower green dashed lines are the predictions by the ANM and the PNM, respectively.}
\label{fig:2}
\end{figure*}

Figures \ref{fig:2}(a) and (b) show $G$ (scaled by the effective strand density $\nu_{\rm eff}$) as a function of $\rho_0$ for the real and phantom networks with strand length $N=10$ and $N=50$, respectively. The modulus of the diamond networks is, as expected, independent of $\rho_0$, and the $G$ of the real networks is smaller than the one of the phantom counterparts. These observations can be rationalized by the fact that in these regular networks entanglements are absent and hence $G$ is determined solely by the configurational entropy, and upon elongation of the strands this entropy changes more rapidly  for the phantom chains than for the real chains. Note that the modulus of a phantom diamond network might be expected to be given by the value predicted by the PNM, but the figure shows that this is not the case. This discrepancy, which depends on the length of the chains, is likely due to the fact that the end-to-end distances of the strands are not given by a Gaussian distribution, thus violating the second assumption of the PNM. However, in the context of Fig. \ref{fig:8}(b) we will see that the values of $G$ of the phantom diamond networks are in excellent agreement with the predictions by the ANT, therefore confirming that the individual strands in these networks are still Gaussian chains.

For the amorphous networks one finds that $G$ increases (basically linearly) with $\rho_0$, in qualitative agreement with experiments\cite{Akagi2013, Nishi2017}, a trend that cannot be reproduced by the ANM or PNM models (green dashed lines). Below we will argue that this $\rho_0$-dependence is directly related to the density dependence of the pre-strain. Furthermore one sees that for this  type of network the $N$-dependence of the phantom networks is relatively modest, confirming the expectation that entanglement effects are not relevant. This is in contrast to the real networks for which the $G$ for the $N=50$ system is significantly higher than the one for the $N=10$, indicating the presence of entanglements. The weakness of entanglement effects for the $N=10$ system makes that the $G$ of the real network is below the one of the phantom network (for the same reason as in the diamond networks), while for the $N=50$ system the effect of these entanglements become so significant that at intermediate and high $\rho_0$ the trend is inverted. The difference $\Delta G\equiv G_{\rm {real}}-G_{\rm {phant}}$ increases continuously with $\rho_0$, implying stronger entanglements at higher densities, in line with textbook knowledge\cite{rubinstein2003polymer} and the results of polymer melts\cite{everaers2004rheology}.  To support these arguments we present in section S3 of the SI a primitive path analysis \cite{everaers2004rheology} of the strands which shows that (i) the $N=10$ and $N=50$ systems are indeed  slightly and strongly entangled, respectively, and (ii) for $N=50$ the strength of the entanglement is an increasing function of $\rho_0$.

\begin{figure*}
\centering
\includegraphics[width=\textwidth]{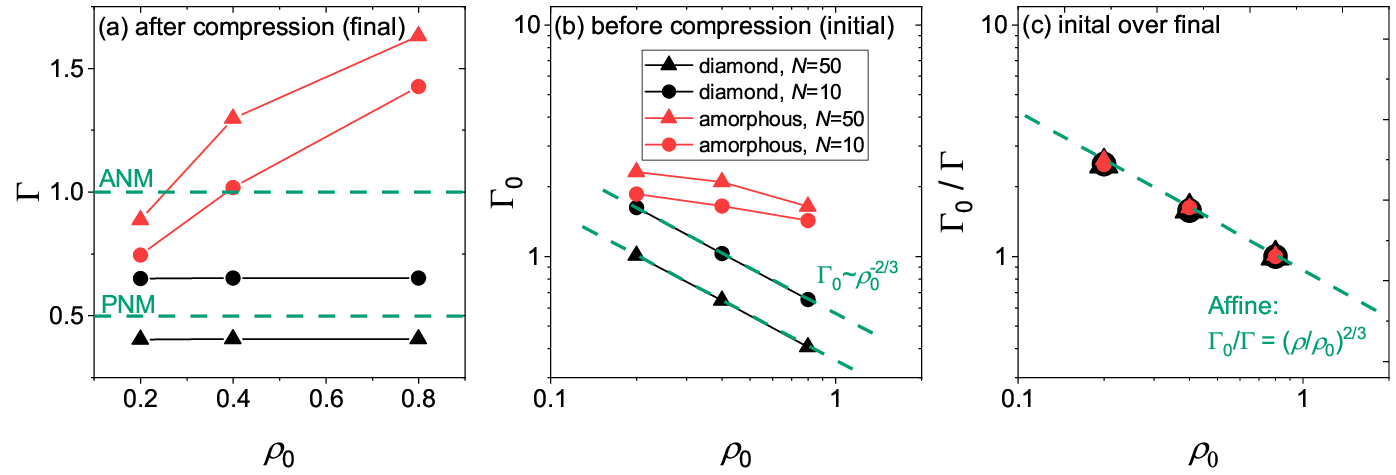}
\caption{Explaining the $\rho_0$ dependence of $G$ of amorphous networks using the pre-strain measure $\Gamma$: (a) $\Gamma$ of final (compressed) networks, (b) $\Gamma_0$ of initial (percolated) networks, and (c) $\Gamma_0/\Gamma$ as a function of $\rho_0$. ANM and PNM refer to the affine network model and phantom network model, respectively.}
\label{fig:3}
\end{figure*}

Figure \ref{fig:2} demonstrates that even if there are no entanglements, such as in the real and phantom amorphous networks with $N=10$ in Fig. \ref{fig:2}(a) or in the phantom amorphous network with $N=50$ in Fig. \ref{fig:2}(b), $G$ is an increasing function of $\rho_0$, implying that other factors than entanglements and defects lead to the observed $\rho_0$ dependence in these networks. 
Since experiments have shown that the pre-strain can rationalize the mechanical properties of networks~\cite{Beech2023} and simulations have demonstrated that ANT is able to predict the $G$ of phantom networks~\cite{Gusev2019numerical,Gusev2019molecular,Tsimouri2020,Gusev2022}, we will in the following probe whether the observed dependencies can be rationalized by means of the pre-strain. To this aim we calculate $\Gamma =\left \langle \frac{\overline{\mathbf{R}} ^{2} }{N_{\rm b}b^2} \right \rangle$ from long-time trajectories to measure how much a network is pre-stretched compared to the Gaussian (melt) state, see discussion in Methods. Figure~\ref{fig:3}(a) shows that $\Gamma$ increases indeed with $\rho_0$ for the amorphous networks but not for the diamond networks, in line with the $\rho_0$ dependence of $G$ shown in Fig. \ref{fig:2}. The $\rho_0$ dependence of $\Gamma$ seen in Fig. \ref{fig:3}(a) provides thus a natural explanation for the $\rho_0$ dependence of $G$ seen in Fig. \ref{fig:2}.

\begin{figure*}
\centering
\includegraphics[width=0.75\textwidth]{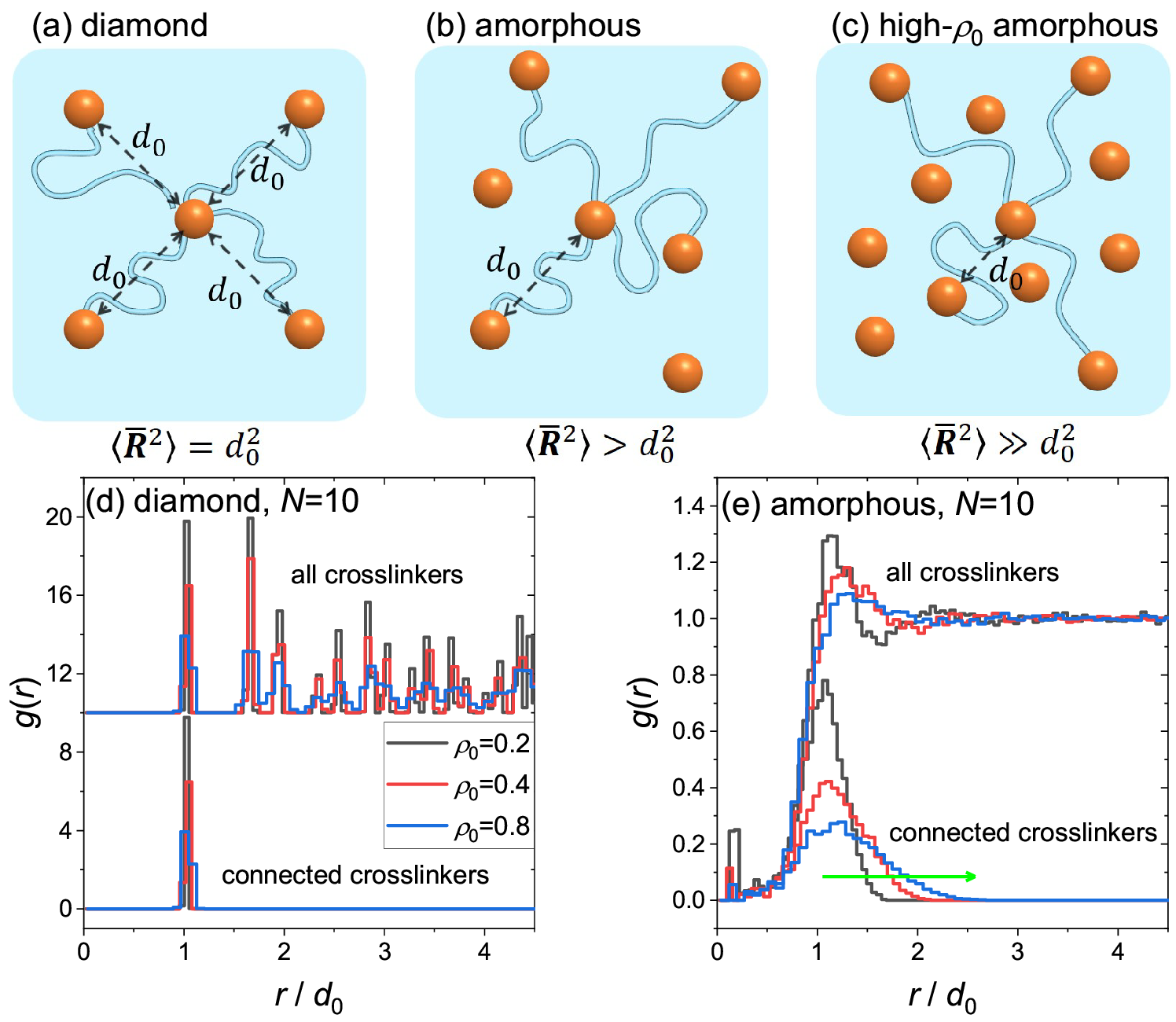}
\caption{Explaining the origin of the $\rho_0$-dependence of $\Gamma$. Top: Sketch of how strands are formed in (a) diamond, (b) amorphous, and (c) high-$\rho_0$ amorphous networks. Bottom: Radial distribution function $g(r)$ of all and connected crosslinkers of initial (d) diamond and (e) amorphous networks (time-averaged configurations) with $N=10$. In (d), curves of all crosslinkers are shifted vertically to avoid overlapping with those of connected crosslinkers. Note that the abscissa variable $r$ is scaled by $d_0$.}
\label{fig:4}
\end{figure*}

To understand the $\rho_0$-dependence of $\Gamma$, we calculate the pre-strain of the networks before the compression, $\Gamma_0$. In Fig. \ref{fig:3}(b) one sees that for both types of networks $\Gamma_0$ decreases with $\rho_0$, which is reasonable, since a smaller initial box (higher $\rho_0$) gives strands that are on average more compressed, i.e., have smaller $\overline{\mathbf{R}}$. For the diamond networks, we find that $\Gamma_0\sim \rho_0^{-2/3}$, which reflects the trivial scaling due to the perfect homogeneity of diamond networks: All strands connect two closest crosslinkers (with distance $d_0$) on a diamond lattice (see Fig. \ref{fig:4}(a)), therefore $\left | \overline{\mathbf{R}} \right |=d_0$. Since for a fixed crosslinker fraction $\rho_0\sim 1/d_0^3$, one has $\Gamma_0\sim \left \langle \overline{\mathbf{R}} ^{2} \right \rangle \sim \rho_0^{-2/3}$. In contrast to the diamond networks, $\Gamma_0$ of the amorphous networks decreases as a function of $\rho_0$ more \textit{slowly} than $\rho_0^{-2/3}$, and below we will rationalize this finding.

Figure \ref{fig:3}(c) demonstrates that upon compression $\Gamma_0/\Gamma$ changes in a nearly affine way, irrespective of the network type, i.e., $\Gamma_0/\Gamma= (\rho/\rho_0)^{2/3}$. While for the diamond network this result is expected, for the amorphous networks it is not trivial since it implies that the statistics of the end-to-end distance of strands does not change under compression. For short chains, where entanglement effects are weak, this is not that surprising, but for the longer chains in the $N=50$ system this is remarkable. For the diamond networks, the density dependence in Fig. \ref{fig:3}(b) and the affine relation in Fig. \ref{fig:3}(c) compensate each other exactly, leading to a $\Gamma$ that is independent of $\rho_0$, in agreement with Fig.~\ref{fig:3}(a). For the amorphous networks, however, the slow decay of $\Gamma_0$ in Fig. \ref{fig:3}(b), cannot counterbalance the quick decay of $\Gamma_0/\Gamma$ in Fig. \ref{fig:3}(c), resulting in an increase of $\Gamma$ with $\rho_0$. Thus one concludes that the $\rho_0$ dependence of $G$ is due to the fact that $\Gamma_0(\rho_0)$ decreases slower than $\rho_0^{-2/3}$. Before discussing this point, we return briefly to the $\rho_0^{-2/3}$-dependence of the data in Fig. \ref{fig:3}(c). Fig. S10 demonstrates that this dependence only holds on the strand-averaged level, while individual strands can experience strong re-organization during the compression, especially for $\rho_0=0.2$ which implies a stronger volume decrease to reach $\rho=0.8$. We also find that the pre-strain of instantaneous configurations does not change according to an affine change of the lengths scales during the compression, see Fig. S11, indicating that time-averaged configurations are more suitable than instantaneous snapshots for the analysis of the pre-strain. 

To rationalize why for the amorphous networks $\Gamma_0(\rho_0)$ decreases only slowly with $\rho_0$, Fig.~\ref{fig:3}(b), we show in Figs.~\ref{fig:4} (a)-(c) cartoons of the networks. The case of the diamond network is shown in panel (a) and one sees that the end-to-end distance of the strands is given by $d_0 \propto \rho_0^{-1/3}$, thus making that $\Gamma_0 \propto \langle \overline{\mathbf{R}}^2 \rangle$ decays like $\rho_0^{-2/3}$. For the amorphous networks, the strands can connect crosslinkers that are beyond $d_0$, making that $\langle \overline{\mathbf{R}}^{2} \rangle > d_0^2$, which explains the vertical ordering of the curves in Fig.~\ref{fig:3}(b), $i.e.$, $\Gamma_0^{\rm {amor}}/\Gamma_0^{\rm {diam}}>1$. This ratio would become larger when $\rho_0$ increases, as a strand can reach more distant crosslinkers, as sketched in Fig. \ref{fig:4}(c). This increasing $\Gamma_0^{\rm {amor}}/\Gamma_0^{\rm {diam}}$ as a function of $\rho_0$ finally makes $\Gamma_0^{\rm {amor}}(\rho_0)$ change more slowly than $\rho_0^{-2/3}$.

To verify this line of argument, we calculate for the initial networks the radial distribution functions $g(r)$ for all the crosslinkers as well as for the connected ones, and present the results for $N=10$ in Fig. \ref{fig:4} (results of $N=50$ are very similar). Figure \ref{fig:4}(d) shows that for the diamond networks all connected crosslinkers come from the closest pairs (the first peak in $g(r)$), in line with Fig. \ref{fig:4}(a). In contrast to this, Fig. \ref{fig:4}(e), one finds that crosslinkers in the higher-$\rho_0$ amorphous networks reach on average more distant neighbors (indicated by the green arrow), supporting the mechanism sketched in Fig. \ref{fig:4}(b) and (c). 

Having verified that the pre-strain measure $\Gamma$ provides a satisfactory qualitative explanation for the observed $\rho_0$ dependence of $G$, we move on to a more quantitative analysis based on $\Gamma$. Before doing this, we briefly discuss the properties of polydisperse networks to explore the universality of the $\rho_0$ dependence of $G$.

\subsection{Polydisperse networks}

\begin{figure*}[th]
\centering
\includegraphics[width=\textwidth]{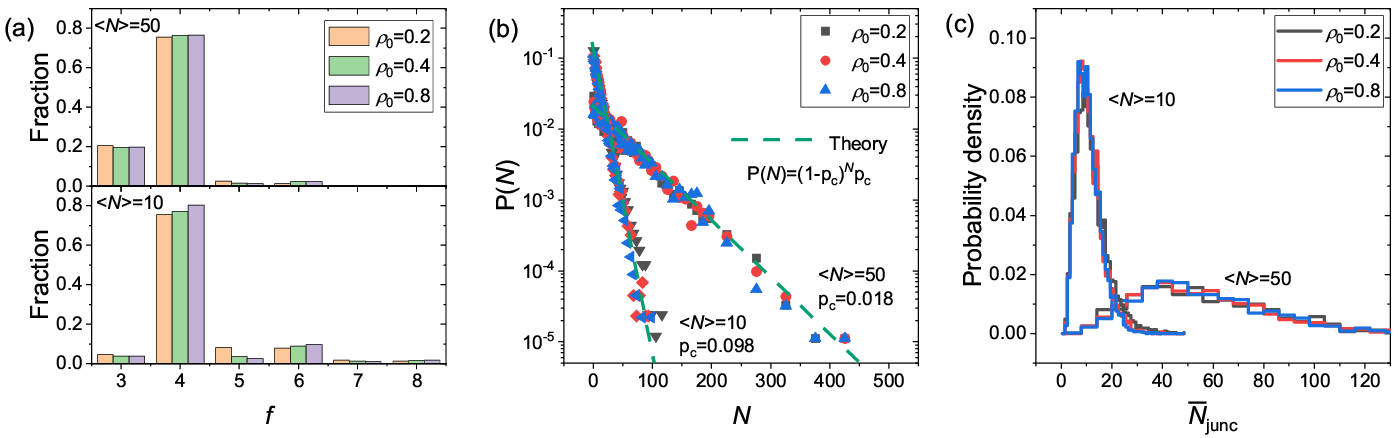}
\caption{Basic structural quantities of polydisperse networks do not depend on $\rho_0$: (a) junction functionality distribution, (b) strand length distribution, and (c) junction-averaged strand length distribution. In (b), $p_{\rm c}$ is the fraction of crosslinkers over all particles.}
\label{fig:5}
\end{figure*}

The nearly defect-free polydisperse networks are generated by randomly adding end-links and cross-links to a polymer solution (see Methods for details). Figure \ref{fig:5} demonstrates that the basic topological quantities are independent of the preparation concentration $\rho_0$. (Here ``topological'' refers to quantities that do not change upon deformation). For example, Fig. \ref{fig:5}(a) shows that the distribution of the functionality of the junctions is not affected by $\rho_0$ in that for all the simulated $\rho_0$, most junctions connect four strands and only a few have less or more strands, in agreement with the results of Gula \textit{et al}\cite{Gula2020}. (The number of junctions with more than eight strands is very small, so we do not show them.) Also the distribution of the strand length, Fig.~\ref{fig:5}(b), is independent of $\rho_0$ and is described well by an exponential, in agreement with previous results\cite{Grest1990,Gula2020,Sorichetti2021}, and with theory (green dashed lines)\cite{Grest1990}. Figure~\ref{fig:5}(c) presents the probability that the average length of the strands that are attached to a given junction, $\overline{N}_{\rm junc}$. This quantity reflects to some extent the constraints felt by the junction. For example, one may expect that a junction with $\overline{N}_{\rm junc}=5$ feels stronger constraints and thus fluctuates less than a junction with $\overline{N}_{\rm junc}=50$. The figure demonstrates that the distribution of $\overline{N}_{\rm junc}$ depends strongly on $N$, but at fixed $N$ it is independent of $\rho_0$, indicating that overall the junctions in the networks with the same $N$ are very similar in terms of connectivity constraints from their strands. 

\begin{figure*}[th]
\centering
\includegraphics[width=0.75\textwidth]{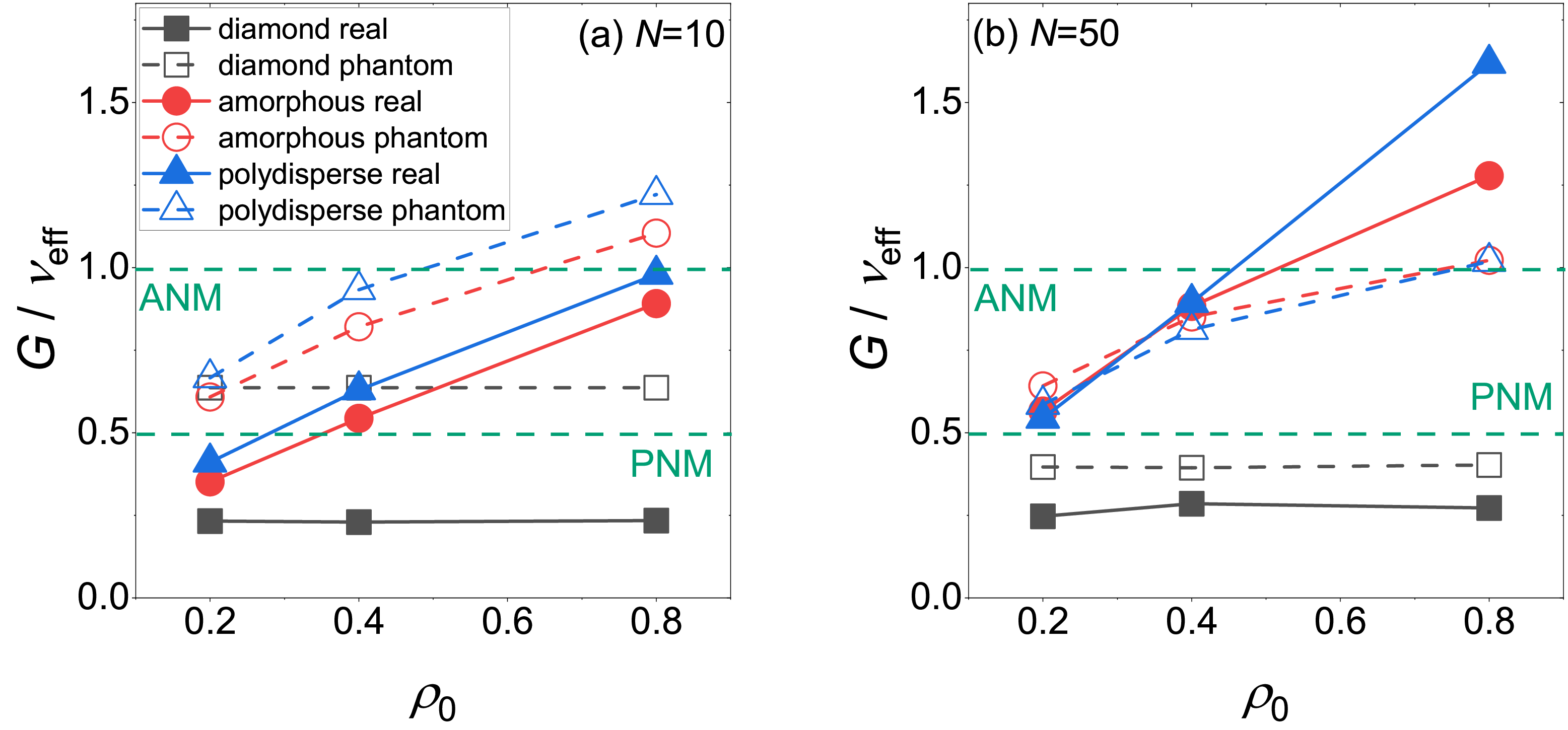}
\caption{Shear modulus $G$ (scaled by $\nu_{\rm eff}$) as a function of preparation concentration $\rho_0$ of simulated monodisperse and polydisperse networks with (a) $\left \langle N \right \rangle=50$ and (b) $\left \langle N \right \rangle=10$. The green dashed lines are the predictions of the affine network model and phantom network model.}
\label{fig:6}
\end{figure*}

Given that from a statistical point of view polydisperse networks generated at different $\rho_0$ have basically identical topology, one might expect their elasticity to be very similar. However, Fig. \ref{fig:6} demonstrates that the polydisperse networks (blue symbols),  show clearly an increase of $G$ as a function of $\rho_0$, just like monodisperse networks which, for the sake of comparison, are included in the graphs as well (red symbols). For the case $N=10$, panel (a), the curves for the two type of network track each other closely, with the $G$ for the polydisperse network a bit above the one of the amorphous system, a ranking that is reasonable since the presence of short chains will render the former network stiffer. This difference is less pronounced for the $N=50$ chains, panel (b), since the presence of very short chains is less likely, see Fig.~\ref{fig:5}(b). For the highest density one does, however, find a strong difference in $G$ between the polydisperse system and the amorphous one. A primitive path analysis indicates that this effect is due to the presence of additional entanglements formed by the very long strands in the polydisperse network, see SI Figs. S4 and S5(a)-(c).

\begin{figure*}
\centering
\includegraphics[width=\textwidth]{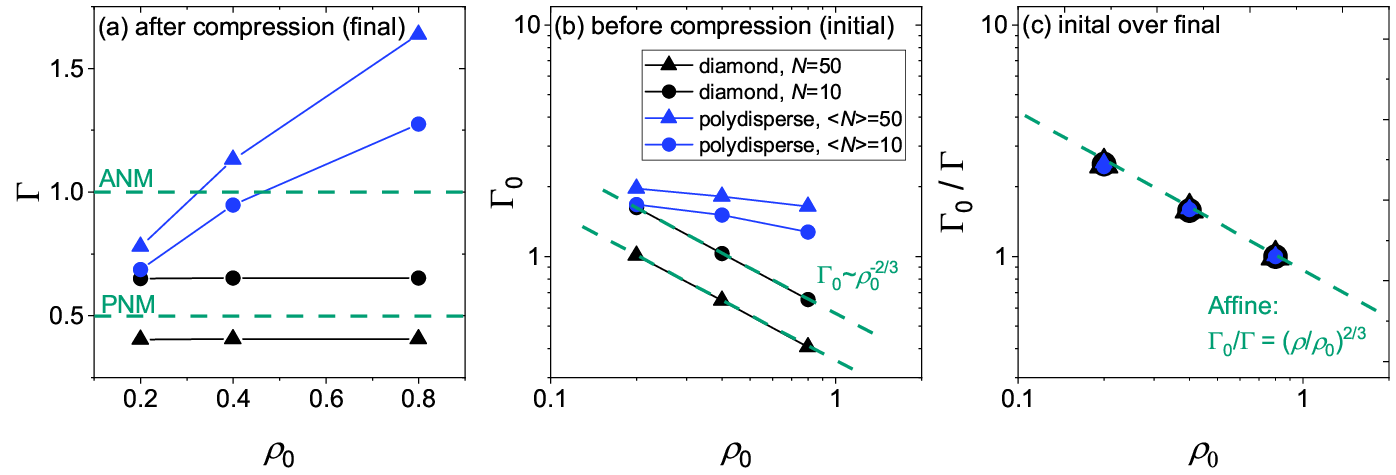}
\caption{Explaining the $\rho_0$-dependence of $G$ of the polydisperse networks using the pre-strain measure $\Gamma$: (a) $\Gamma$ of final (compressed) networks, (b) $\Gamma_0$ of initial (percolated) networks, and (c) $\Gamma_0/\Gamma$ as a function of $\rho_0$. The green dashed lines are the predictions of the affine network model and phantom network model.} 
\label{fig:7}
\end{figure*}

Figure~\ref{fig:7} parallels Fig.~\ref{fig:3} and shows that the $\rho_0$-dependence of $G$ for the polydisperse networks can be explained by the pre-strain measure $\Gamma$, as it was the case for the amorphous networks. Similar to amorphous networks, Fig. \ref{fig:7}(c) shows that for the polydispers networks $\Gamma$ changes in an affine way during the compression from $\rho_0$ to $\rho$. Hence the variation of $\Gamma_0$ shown in Fig~\ref{fig:7}(b), weaker than $\rho_0^{-2/3}$, is at the origin of the $\rho_0$ dependence of $\Gamma$ in Fig. \ref{fig:7}(a). Compared to the amorphous networks generated via the reaction of star polymers, the polydisperse networks are generated in a simpler way, $i.e.$, randomly and instantaneously adding links to the close pairs in a precursor (single-stranded) polymer solution. In this way each pre-existing chain is suddenly divided into subchains having an exponential distribution for the strand length $N$.

To estimate the end-to-end distance of the so-generated strands we assume them to be free polymer chains. We recall that $R(N_{\rm b})\sim N_{\rm b}^{x}$\cite{rubinstein2003polymer}, where the Flory exponent $x$ is $1/2$ for a melt ($\rho_0=0.85$) and $3/5$ for a dilute system, $i.e.$, $\rho_0<\rho^*$ where $\rho^*$ is the overlap concentration. We infer from this that  $\Gamma_0(\rho_0<\rho^*) / \Gamma_0(\rho_0=0.85) \simeq N_{\rm b}^{1/5}$, which implies that the change in $\Gamma_0$ is weak when we go from low to high density for the average strand length considered in the present study, i.e., $50^{1/5}=2.19$ and $10^{1/5}=1.58$. Note that both of these values are smaller than the effect of the affine scaling between $\rho_0=0.2$ and $\rho_0=0.8$, $i.e.$ $(0.2/0.8)^{-2/3}=2.52$. Networks generated from concentrations between the two limits can thus be expected to show a slow variation of $\Gamma_0$ with $\rho_0$, in agreement with Fig. \ref{fig:7}(b). Hence the weak variation of $\Gamma_0$ seen in Fig. \ref{fig:7}(b) appears to be a direct result of randomly crosslinking the pre-existing random-coil-like chains, and we can expect that the $\rho_0$ dependence of $G$ will be a quite general phenomenon in elastomer synthesis.

One may question the above analysis by pointing out that the Flory exponent only applies to long sub-chains while our polydisperse networks are dominated by short strands. However, according to the blob-scaling theory\cite{de1979scaling,rubinstein2003polymer}, short sub-chains only detect the local packing environment and thus behave like either ideal chains (when below the thermal blob length scale) or self-avoiding chains (when below the correlation blob length scale), neither of which gives a $\rho_0$-dependent $R$. As a result, $\overline{\mathbf{R}}$ and also $\frac{\overline{\mathbf{R}} ^{2} }{N_{\rm b}b^2}$ of short strands should show much weaker $\rho_0$ dependence than that of long strands, which is verified by our simulated data in Fig. S12 in the SI.

\subsection{Quantitative application of ANT}

\begin{figure*}
\centering
\includegraphics[width=\textwidth]{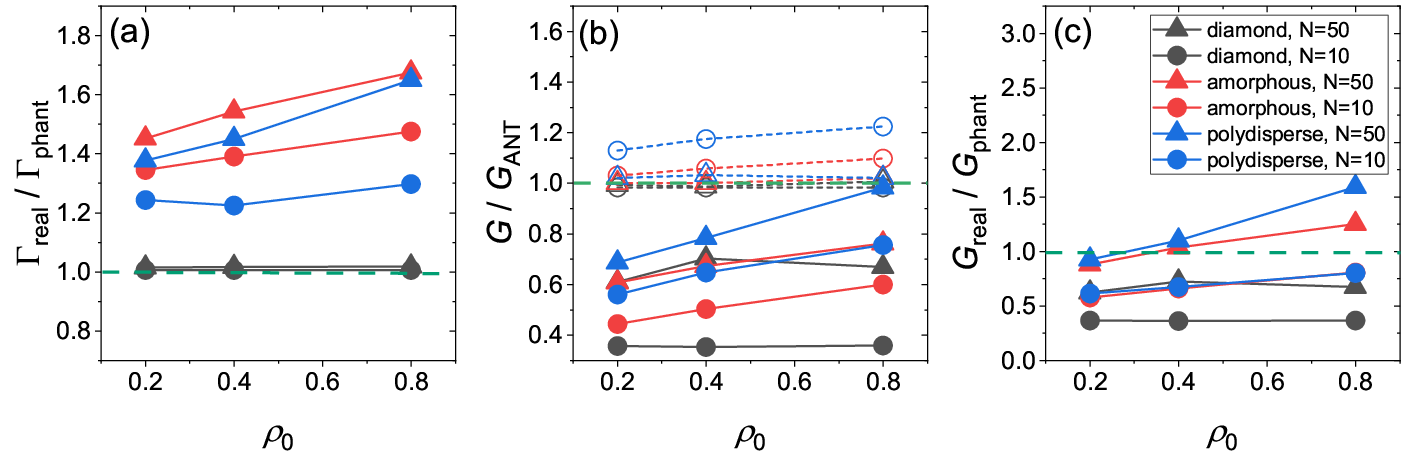}
\caption{Testing the prediction of the ANT regarding the simulated networks: (a) $\Gamma_{\rm real}/\Gamma_{\rm phant}$, (b) $G/G_{\rm ANT}$, and (c) $G_{\rm real}/G_{\rm phant}$ as a function of $\rho_0$. In (b), full symbols with solid lines correspond to real networks, while open symbols with dashed lines correspond to phantom networks. In all panels, the green dashed line indicates the value of the ordinate 1.0.}
\label{fig:8}
\end{figure*}

Since we have seen that the pre-strain measure $\Gamma$ allows to explain qualitatively the $\rho_0$-dependence of $G$, it is attempting to test whether this quantity is able to give also a quantitative prediction of $G$. Since the relation predicted by the ANT, $G_{\rm ANT}=\Gamma\nu k_{\rm B}T$, has been shown to be able to predicts the elastic moduli of ideal networks made up of monodisperse phantom Gaussian strands~\cite{Gusev2019numerical,Gusev2019molecular,Tsimouri2020}, we test in the following whether this is also the case for the networks considered in the present work. For the simulated real networks, $\Gamma(\rho_0)$ has already been presented in Figs. \ref{fig:3}(a) and \ref{fig:7}(a). We calculate $\Gamma$ for the corresponding phantom networks, and show in Fig.~\ref{fig:8}(a) the ratio $\Gamma_{\rm real}/\Gamma_{\rm phant}$ as a function of $\rho_0$, where the subscripts ``real'' and ``phant'' denote the real and phantom networks, respectively. For the diamond networks one finds, as expected, that $\Gamma_{\rm real}=\Gamma_{\rm phant}$,  since real and phantom diamond networks share the same time-averaged configuration with crosslinkers localize at the sites of the diamond lattice, see the snapshot in Fig. \ref{fig:1}(c). For the two disordered networks one notes instead that $\Gamma_{\rm real} > \Gamma_{\rm phant}$, i.e., the pre-strain decreases if volume exclusion is removed. This is reasonable since the local constraints in the real networks can be released when chain-crossing are allowed,  leaving only the global-level constraints associated with the crosslinks. Note that this decrease is expected to be more pronounced if the strands are long, since there are more entanglements, in agreement with the observation that the curves for $N=50$ are above the ones for $N=10$. The graph also shows that the ratio $\Gamma_{\rm real}/\Gamma_{\rm phant}$ increases slowly with $\rho_0$, a trend that can be rationalized by recalling that the pre-strain of the real network increases with $\rho_0$ and hence more constraints are released when it is transformed into its phantom counterpart.

In Fig. \ref{fig:8}(b) we plot $G/G_{\rm ANT}$ as a function of $\rho_0$ for both real (full symbols with solid lines) and phantom (open symbols with dashed lines) networks, where $G$ is the modulus obtained from the simulations, see Fig. \ref{fig:6}, and $G_{\rm ANT}$ is the modulus predicted by Eq.~\ref{eq:6}. While Gusev and co-authors\cite{Gusev2019numerical,Gusev2019molecular,Tsimouri2020} have applied the ANT to monodisperse amorphous phantom networks generated by a Monte Carlo method, the applicability of the ANT for diamond and polydisperse networks has not yet been investigated. Figure~\ref{fig:8}(b) reveals that most moduli of the simulated phantom networks are well predicted by the ANT (green dashed line), and even for the two worst cases (red and blue open circles) the deviations are less than 20\%, see also discussion below. Since the networks generated with different $\rho_0$ but identical $\left \langle N \right \rangle$ have nearly the same strand density, the result $G=G_{\rm ANT}$ implies that the increase of $G$ with $\rho_0$ for the phantom networks originates solely from the increase of $\Gamma_{\rm phant}$. In other words, Fig. \ref{fig:8}(b) shows that the $\rho_0$ dependence of $G$ for phantom networks can be fully $quantitatively$ explained by the pre-strain quantity $\Gamma_{\rm phant}$ based on the ANT.

The phantom networks with the largest deviations from the ANT prediction in Fig. \ref{fig:8}(b) are the amorphous and the polydisperse networks with $N=10$. For these short strands, the slightly higher values of $G$ are probably induced by the non-Gaussian extensions of the strands, as a strand stretched beyond the Gaussian regime will provide a force higher than the harmonic prediction\cite{Sorichetti2021}. The deviation of $G/G_{\rm ANT}$ from 1 for the $N=10$ disordered phantom networks is slightly stronger when $\rho_0$ is higher, probably because the networks with higher $\rho_0$ are more pre-stretched and thus their short strands enter the non-Gaussian extension regime more easily. Note that also the polydisperse networks with $N=50$ possess a fraction of very short strands including $N=1$ and $N=2$ (see Fig. \ref{fig:5}(b)), but these networks have moduli (open blue triangles) close to the ANT estimate (with a deviation smaller than 5\%), which is a little surprising as the ANT is based on the Gaussian-extension assumption which usually does not apply to very short strands. This surprising success of ANT to the phantom polydisperse networks with $N=50$ raises the necessity of examining in detail how the very short strands behave upon deformation of the whole network, i.e., whether the presence of small regions with high stiffness is at the end irrelevant for the mechanical deformation of the network. Addressing this question is important but beyond the scope of the present paper and hence should be the subject of future investigations.

The results for the simulated real networks are shown in Fig. \ref{fig:8}(b) as full symbols with solid lines. For most networks one recognizes that $G$ differs from $G_{\rm ANT}$, which is not surprising, since the derivation of the ANT is based on the assumption of phantom Gaussian strands and affine movements, neither of which is satisfied by the real networks (Gusev and Schwarz\cite{Gusev2022} showed that the affine assumption does not hold for real networks). Furthermore we find that $G<G_{\rm ANT}$ for all the simulated real networks. This can be understood from the fact that the ANT assumes phantom strands whose motions are not limited by steric hindrance, while in a real network volume exclusion leads to a reduction of the accessible configuration space, and hence of the entropy change upon deformation, which results in less stress and hence a smaller modulus.

Given that $G_{\rm phant}$, the modulus of a phantom network, can be estimated easily via the ANT combined with the maximum-entropy procedure\cite{Gusev2019numerical,Gusev2019molecular,Tsimouri2020}, it is interesting to compare the modulus of a real network, $G_{\rm real}$, also with $G_{\rm phant}$. Actually, as shown by Gusev and Schwarz\cite{Gusev2022} for end-linked monodisperse networks and by Gula et al.\cite{Gula2020} for randomly-crosslinked polydisperse networks, simply adding the entanglement modulus $G_{\rm e}$ of the precursor melt to the modulus of the phantom network gives quite good estimates for the real networks generated from a melt, i.e., $G_{\rm real} \approx G_{\rm phant}+G_{\rm e}$, and hence this method is widely used in the literature to estimate the modulus of a real network, given that $G_{\rm e}$ has abundant experimental data or can be easily estimated from simulations of melts via the primitive path analysis. However, this approximation is obviously not reliable for the case of the diamond networks and the $N=10$ disordered (amorphous and polydisperse) networks in the present study, as these networks have moduli $lower$ than those of their phantom counterparts (see Fig. \ref{fig:6}), implying that $G_{\rm e}$ has an unphysical negative value.

To compare $G_{\rm real}$ with $G_{\rm phant}$ we therefore consider the ratio $G_{\rm real}/G_{\rm phant}$, which is displayed in Fig. \ref{fig:8}(c). One sees that for the diamond and the $N=10$ disordered networks one has $G_{\rm real}/G_{\rm phant}<1$, underscoring the effect of decreasing entropy due to volume exclusion. For the disordered $N=10$ networks, $G_{\rm real}/G_{\rm phant}$ increases with $\rho_0$, which is expected due to the stronger local constraints at higher $\rho_0$ that increase the pre-stretch, as already evoked in the increasing trend of $\Gamma_{\rm real}/\Gamma_{\rm phant}$ in Fig. \ref{fig:8}(a). Interestingly, the curves of the amorphous and the polydisperse networks with $N=10$ (red and blue circles) almost overlap with each other, indicating that, despite the greatly different network architectures, the volume exclusion affect the moduli of small-$\left \langle N \right \rangle$ amorphous and polydisperse networks in a very similar way. For the $N=50$ disordered networks, due to the stronger effect of entanglements, $G_{\rm real}$ begins to surpass $G_{\rm phant}$ for $\rho_0 \gtrsim 0.4$, especially for the polydisperse one which have very long strands and thus more entanglements, similar to what we have seen in Fig.~\ref{fig:6}.

The above discussion leads thus to the conclusion that the influence of the volume exclusion on the network elasticity is related to three (not necessarily independent) factors: (1) Homogeneous reduction of the accessible configuration space; (2) Increase of local constraints resulting in an enhancement of the pre-stretch (without changing the statistics of the strand numbers); (3) Introduction of entanglements and thus increase of the (effective) strand number (and therefore density). The first factor decreases the modulus, while the latter two increase it, and the final value of the modulus is determined by the combination of the three factors. The diamond networks are only affected by the first factor, the $N=10$ disordered networks include the first two, while the $N=50$ disordered networks involve all the three points. As far as we know, while in the past the aspect of entanglements has been  analyzed in detail, the other two have received less attention. Therefore, our results highlight the limitations of the current theoretical understanding of rubber elasticity and calls for approaches in which the role of volume exclusion is precisely and comprehensively analyzed.

\section{Conclusions and outlook}

Using molecular dynamics simulations of coarse-grained polymer models, we have investigated how the structural and mechanical properties of polymer networks depend on their production history and their architecture. In particular we have considered how the initial concentration $\rho_0$ of the monomers, the length of the strands, as well as the excluded volume influence the elastic modulus $G$ of the sample. For the regular diamond networks we find that $G$ is independent of $\rho_0$, while for disordered (amorphous and polydisperse) networks $G$ shows a clear increase with $\rho_0$, results that are in line with experimental results and hence is evidence that our simulations reflect qualitatively the behavior of real systems. The observed $\rho_0$-dependence of $G$ for the three types of networks can be rationalized by means of the pre-strain of the networks, which in turn can be easily measured by the quantity $\Gamma$ introduced within the framework of the ANT, thus underscoring the relevance of the pre-strain to understand the mechanical properties of polymer networks. Furthermore we demonstrate that the ANT makes a reliable prediction for the elasticity of the phantom networks but fails to do so for the case of real networks, and trace back this deficiency to the effect of volume exclusion and entanglements.

Our work documents several interesting phenomena that deserve further explorations. Our results, including Fig. \ref{fig:6}, Fig. \ref{fig:8}(b) and (c), consistently show that amorphous and polydisperse networks with the same average $N$ and $\rho_0$ have a very similar elastic behavior even if their strand length distributions are quite different. This similarity is surprising, as very short strands are abundant in polydisperse networks and can be expected to extend upon deformation in a non-Gaussian way~\cite{Sorichetti2021}, quite different from the random-coil-like strands in monodisperse networks. Therefore it would be valuable to observe in detail how a polydisperse network re-organizes its structure under deformation and how this behavior changes as $\left \langle N \right \rangle$ increases. It will thus be interesting to investigate further why the short and thus stiff strands do not affect in a significant manner the elastic properties of the network and how this observation depends on $\left \langle N \right \rangle$, the average length of the strands.

A further notable phenomenon observed in the present simulations, documented in Fig.~\ref{fig:8}(c), is that for the amorphous and polydisperse systems with $N=50$ the ratio $G_{\rm {real}}/G_{\rm {phant}}$ exceeds 1.0 at around $\rho_0=0.4$. The same transition is observed at $\rho_0 \approx 0.4$ when $N$ increases from 10 (circles) to 50 (triangles) for either amorphous or polydisperse. A common (implicit) belief in polymer physics is that volume exclusion introduces entanglements into the system and thus increases the elastic modulus. However, here one sees that when the strands are not strongly entangled (as it is the case in the diamond networks) or when the strands are short (like the $N=10$ systems), the volume exclusion can {\it decrease} the modulus, i.e.,  $G_{\rm {real}}/G_{\rm {phant}}<1$. Uncovering and describing on a quantitative level the details that lead to this reduction of the elasticity would certainly improve significantly our understanding of entanglement effects in polymer networks and hence such a study should be done in the future.

%% classes. This should be given within the "acknowledgement"
\begin{acknowledgement}

We thank M.~Le Goff, M.~Bouzid, and K.~Martens for useful discussions. JT acknowledges the financial support by the China Scholarship Council (CSC) (No. 201904890006) and by the National Natural Science Foundation of China (No. 52303029). Part of the work was done when JT visited LIPhy in Grenoble as a CSC postdoc, so JT appreciates the colleagues at LIPhy for the welcoming environment and the active scientific atmosphere. JT is also grateful to his employer INPC for the strong support which makes his stay in Grenoble possible. WK is a senior member of the Institut Universitaire de France.

\end{acknowledgement}

\begin{suppinfo}

Supporting information is available at:

\end{suppinfo}


\bibliography{library_abbreviated}

\end{document}